\documentclass[num,preprint,review,12pt]{elsarticle}

\usepackage{amssymb}
\usepackage{amsthm}
\usepackage{amsmath}
\usepackage{subcaption}
\usepackage{algorithm}
\usepackage{algpseudocode}

\newcommand{\reals}{{\mbox{\bf R}}}
\newcommand{\dom}{\mathop{\bf dom}}
\newcommand{\rint}{\mathop{\bf ri}}

\newcommand{\Prob}{\mathop{\bf Prob}}

\newcommand{\ExpFt}[1]{\mathop{\mathbb{E}{}}\left[#1\mid\mathcal{F}_t\right]}

\newtheorem{theorem}{Theorem}[section]

\journal{Signal Processing}

\begin{document}

\begin{frontmatter}

\title{StochaLM: a Stochastic alternate Linearization Method for distributed optimization}

\author[IST,ISR]{In{\^e}s Almeida\corref{cor1}\fnref{bolsas}}
\ead{[last].[first]b@gmail.com}
\cortext[cor1]{Corresponding author.}
\fntext[bolsas]{Work partially funded by Funda{\c c}{\~a}o para a Ci{\^e}ncia e Tecnologia (FCT), Portugal, under grants PD/BD/135012/2017 and BL40/2021.}

\author[IST,ISR]{Jo{\~a}o Xavier\fnref{projectos}}
\fntext[projectos]{Work partially funded by FCT, Portugal, under Project UID/EEA/50009/2013.}

\affiliation[IST]{organization={Instituto Superior T{\'e}cnico},
            city={Lisbon},
            country={Portugal}}

\affiliation[ISR]{organization={Institute for Systems and Robotics},
            city={Lisbon},
            country={Portugal}}

\begin{abstract} 
We present the Stochastic alternate Linearization Method (StochaLM), a token-based method for distributed optimization. This algorithm finds the solution of a consensus optimization problem by solving a sequence of subproblems where some components of the cost are linearized around specific anchor points. StochaLM can be interpreted as a dual block coordinate ascent method whose block components are selected using the state of an ergodic Markov chain. This sampling process is neither essentially cyclic nor independent over time, preventing us from using proofs of convergence of dual block coordinate ascent methods done in previous works. The proof of convergence of our method is, therefore, also novel. We show that, if the cost is strongly convex and the network is fully connected, then, with probability one, the primal sequence generated by StochaLM converges to the solution. Our method is application-friendly, as it has no global hyperparameters to tune; any hyperparameters can be tuned locally by the agents, using information regarding their private cost and neighbourhood. Our method is, therefore, decentralized in the truest sense. Numerical experiments evidence that our method converges to the solution faster than other token-based methods, even when these methods’ hyperparameters are tuned for optimal performance.
\end{abstract}


%
%
%


\end{frontmatter}


\section{Introduction}\label{sec:intro}

More and more, data is being collected in a distributed way: Networks of sensors take local measurements of a system, teams of robots move in hazardous environments while tracking a target, and smartphones collect their users' daily activity. In a traditional, centralized approach, a central machine collects all the data, and then performs the required computations to extract useful information. In a distributed approach, the agents of the network cooperate to reproduce the centralized solution.

Consider $n$ nodes (or agents) which are the vertices of a connected graph. The edges of the graph correspond to valid communication channels between pairs of agents. The agents can communicate information, but not the dataset itself, with their immediate neighbours. Our goal is to solve
\begin{equation} \label{prob:main}
\min_{x\in\reals^p}\ F(x)\equiv f_0(x) + \sum_{i=1}^n f_i(x),
\end{equation}
where $f_0:\reals^p\rightarrow \reals\cup\{+\infty\}$ is a closed, strongly convex function, and the $f_i:\reals^p\rightarrow \reals\cup\{+\infty\}$, $i=1,\ldots,n$, are closed, convex functions. One can think of each $f_i$ within the sum as being the loss on the private dataset stored in the $i$-th node of the network, while $f_0$ represents a data-independent regularizer that is known by all the nodes. Problem~\eqref{prob:main} has a unique (centralized) solution, $x^*:=\arg\min F(x).$

Problem~\eqref{prob:main} finds application in distributed problems such as LASSO regression~\cite{mateos2010distlinreg}, SVMs~\cite{forero2010distsvms}, RVFL networks~\cite{scardapane2015rvfl}, spectrum sensing~\cite{bazerque2010spectrumsensing}, resource allocation~\cite{xiao2004routing,xiao2006resourcealloc}, model predictive control~\cite{panagiotis2013control}, and economic dispatch~\cite{yang2017dispatch}.

\paragraph{Original contributions} We present the Stochastic alternate Linearization Method (StochaLM), a token-based distributed optimization method. This method finds the solution of \eqref{prob:main} by solving a sequence of subproblems where most of the cost components are approximated by linear functions. In the dual domain, StochaLM can be interpreted as a distributed implementation of the stochastic dual block coordinate ascent (DBCA) method where the dual block components are selected according to the state of an ergodic Markov chain. As far as we are aware, the DBCA method has not been adapted to distributed setups before the present work.

The selection of the blocks in StochaLM is neither essentially cyclic, as assumed in~\cite{tseng1993dbca}, nor independent over time, as assumed in~\cite{shwartz2013sdca}. This prevents us from using the theoretical results from those works to study StochaLM; our theoretical results presented here are novel in themselves. We are able to show that, if the cost is strongly convex and the network is fully connected, the accumulation points of the dual sequence generated by our method are optimal, and that the primal sequence converges to the centralized solution.


The main advantage of our method, when compared to other token-based distributed methods, is that it does not have global hyperparameters: Any parameters which may be necessary to fully implement this method can be set locally. For instance, each agent may choose an algorithm to solve its internal subproblems; setting hyperparameters of this algorithm requires, at most, knowledge regarding that agent's private cost (and $f_0$). Also, the transition probabilities that determine the token's movement can be set locally, using only information regarding the agent's immediate neighbourhood, as shown in~\cite{johansson2010token}. These characteristics allow our method to be implemented in a truly decentralized way, making it application-friendly.

We compare our method with other token-based methods by means of numerical simulations. The results suggest that StochaLM converges more quickly to the centralized solution than other token-based methods, even when the (global) hyperparameters of these other methods are tuned for optimal performance.

\paragraph{Document structure} We present StochaLM in Section~\ref{sec:algorithm}, and related work in Section~\ref{sec:related}. The theoretical guarantees of our method are presented and proved in Section~\ref{sec:stochalm_proof}. Numerical results are reported in Sectionç~\ref{sec:numerical}, and final remarks are done in Section~\ref{sec:conclusions}.

\section{Algorithm}\label{sec:algorithm}

We begin by presenting StochaLM in its original, primal form, as this makes the transition from centralized to distributed implementations straightforward. Later in this section, we show that our method can be interpreted as an instance of the dual block coordinate ascent (DBCA) method. This interpretation allows for a better understanding of how StochaLM relates to existing literature, and it greatly facilitates our theoretical analysis.



\subsection{Centralized StochaLM}

Consider that there are $n+1$ vectors of size $p$, which are jointly represented as $\mathbf{x}=(x_0,x_1,\ldots,x_n)\in\reals^{(n+1) p}$. The use of this many copies of the optimization variable will allow us to seamlessly move to the distributed version of this algorithm. In addition, we also have $n$ vectors of size $p$, $g_1,\ldots,g_n$, which store certain subgradients of their respective $f_i$.

The method is initialized by setting $x_i^{(0)} \in \arg\min f_i(x_i)$, and $g_i^{(0)}\in \partial f_i\left(x_i^{(0)}\right)$, for $i=1,\ldots,n$. By construction, one may choose $g_i^{(0)}=0$.

At round $t$, pick some $j(t)\in\{1,\ldots,n\}$; we will specify the way the $j(t)$ are chosen in the next section. Given $j(t)$, the components $f_{j(t)}$ and $f_0$ are the ones which are \emph{not} approximated in this round. The remaining functions are replaced by their linear approximations around some previously computed $x_i^{(t-1)}$. The next step is to solve
\begin{equation} \label{prob:submain}
x^{(t)} = \arg\min_x\ \left\{f^{(t)}(x) \equiv f_0(x) + f_{j(t)}(x) + \sum_{i\neq j(t)} \hat{f}_i^{(t-1)}(x)\right\},
\end{equation}
where $\hat{f}_i^{(t-1)}$ is the linearization of $f_i$ around $x_i^{(t-1)}$,
\begin{equation}\label{eq:f_hat}
\hat{f}_i^{(t-1)}(x) := f_i\left(x_i^{(t-1)}\right) + \left\langle g_{i}^{(t-1)}, x - x_{i}^{(t-1)}\right\rangle.
\end{equation}
Each $g_i^{(t-1)}\in\partial f_i\left(x_i^{(t-1)}\right)$ is a subgradient of $f_i$ at $x_i^{(t-1)}$ that satisfies the optimality condition which we will specify in a moment. We now set
\begin{equation}\label{eq:update_xi}
x_i^{(t)}=\left\{\begin{matrix}
x^{(t)} & \mbox{if } i = j(t), \\
x_i^{(t-1)} & \mbox{otherwise.} \\
\end{matrix}\right.
\end{equation}
Notice that $x^{(t)}$ is unique by the strong convexity of $f_0$. Additionally, we set
\begin{equation}\label{eq:update_gjt}
g_i^{(t)}\in\left\{g_i\in \partial f_i\left(x_i^{(t)}\right)\mid 0\in \partial f_0\left(x_i^{(t)}\right) + g_i + \sum_{k\neq i} g_k^{(t-1)}\right\},
\end{equation}
if $i=j(t)$, and $g_i^{(t)}=g_i^{(t-1)}$ otherwise. In words, $g_{j(t)}^{(t)}$ is a subgradient of $f_{j(t)}$ at $x_{j(t)}^{(t)}$ which satisfies the first order condition for optimality of $f^{(t)}$,
\begin{equation} \label{eq:optimality}
0\in\partial f_0\left(x_{j(t)}^{(t)}\right) + g_{j(t)}^{(t)} + \sum_{i\neq j(t)}  g_i^{(t-1)}. 
\end{equation} 

Finally, since $x_{j(t)}^{(t)}$ and $g_{j(t)}^{(t)}$ are defined, one can compute a (new) linearization of $f_{j(t)}$, $\hat{f}_{j(t)}^{(t)}$, to be used in future rounds. The method continues by picking a new $j(t+1)$, and repeating the steps described above.


\subsection{Distributed StochaLM}

To move to the distributed version of StochaLM, one begins by re-writting problem~\eqref{prob:main} as the equivalent distributed consensus problem
\begin{align}\label{eq:main_dist_vector}
\min_{\mathbf{x}} &\ f(\mathbf{x}) \equiv f_0(x_0) + \sum_{i=1}^n f_i(x_i) \\
\mbox{s.t.} &\ \mathcal{A}\mathbf{x}= 0, \nonumber
\end{align}
where $\mathcal{A}$ is the $np$-by-$(n+1)p$ block matrix given by
\begin{equation}\label{eq:mathcal_A}
\mathcal{A}:=
\begin{bmatrix}
I & -I & 0 & \cdots & 0 \\
I & 0 & -I & \cdots & 0 \\
\vdots & \vdots & \vdots & \ddots & \vdots \\
I & 0 & 0 & \cdots & -I 
\end{bmatrix}.
\end{equation}

The choice of variables, $\mathbf{x}=(x_0,x_1,\ldots,x_n)$, done in the previous section, now gains meaning: For $i=1,\ldots,n$, $x_i$ is the estimate of the solution of problem~\eqref{prob:main} held by agent $i$. As we shall soon see, $x_0$, which links all the other variables together, has physical meaning as well.

Consider now the subproblem solved at each round $t$, that is, problem~\eqref{prob:submain}. If one removes constant terms, one finds that
\begin{equation}\label{eq:submain_agent}
x^{(t)}=\arg\min_x\ f_0(x) + f_{j(t)}(x) + \sum_{i\neq j(t)} \left\langle g_{i}^{(t-1)}, x \right\rangle.
\end{equation}

In a distributed setup, and at the beginning of round $t$, agent $i$ knows its (private) $f_i$, the data-independent regularizer $f_0$, and all the $g_{i}^{(\tau)}$ for $\tau<t$. As such, if agent $j(t)$ were to know the value of $\bar{g}^{(t-1)} := \sum_{i=1}^n g_{i}^{(t-1)}$, then it could solve problem~\eqref{eq:submain_agent}, since $\sum_{i\neq j(t)} g_{i}^{(t-1)} = \bar{g}^{(t-1)} - g_{j(t)}^{(t-1)}$.

Implementing StochaLM in a distributed fashion boils down, therefore, to ensuring that $\bar{g}^{(t-1)}$ is known by agent $j(t)$ at the beginning of round $t$. We describe how this can be done in practice in the following paragraphs.

Consider a token that can travel across the network. This token is initialized by the agent holding it with some value of $x_0^{(0)}$, and $\bar{g}^{(0)}=0$. The agents initialize their own internal variables, $x_i^{(0)}$ and $g_i^{(0)}$, the same way as in the centralized setup.

The procedure for passing the token around can be described as follows: Consider an ergodic Markov chain with $n$ states, where the $i$-th state corresponds to the token being on the $i$-th agent of the network. The transition probabilities of this Markov chain are chosen so that the probability of transitioning from state $i$ to state $j$ is zero if there is no edge linking agent $i$ to agent $j$. Given that the token is in some agent $j(t-1)$, the token will move to one of this agent's neighbours, $j(t)\in\mathcal{N}_{j(t-1)}$, according to these transition probabilities.

Upon receiving the token at round $t$, node $j(t)$ uses the value of $\bar{g}^{(t-1)}$ stored in the token to solve~\eqref{eq:submain_agent}. This node updates its internal variables using~\eqref{eq:update_xi} and~\eqref{eq:update_gjt}, and it also updates the token variables, by setting $x_0^{(t)}=x_{j(t)}^{(t)}$, and by updating $ \bar{g}^{(t-1)}$ to $\bar{g}^{(t)}$ via $\bar{g}^{(t)} = \bar{g}^{(t-1)} - g_{j(t)}^{(t-1)} + g_{j(t)}^{(t)}$. The other agents update their internal variables using~\eqref{eq:update_xi} and $g_i^{(t)}=g_i^{(t-1)}$.

In the beginning of the next round, agent $j(t)$ picks one of its immediate neighbours to pass the token to (using the transition probabilities). This neighbour becomes node $j(t+1)\in\mathcal{N}_{j(t)}$, and the algorithm continues.


\subsection{Dual perspective}\label{sec:stochalm_dual}

Suppose we try to solve the dual problem of~\eqref{eq:main_dist_vector},
\begin{equation}\label{eq:main_dual}
\min_{\mathbf{p}} h(\mathbf{p}) \equiv f^*_0\left(\sum_i p_i\right) + \sum_{i=1}^n f^*_i(-p_i),
\end{equation}
with dual variable $\mathbf{p}=(p_1,\ldots,p_n)\in\reals^{np}$, using block coordinate descent:
\[
p_i^{(t)}\in \arg\min_p\ h\left(p^{(t-1)}_1,\ldots,p^{(t-1)}_{i-1},p,p^{(t-1)}_{i+1},\ldots,p^{(t-1)}_n\right)
\]
if $i = j(t)$, and $p_i^{(t)}=p_i^{(t-1)}$ otherwise. It can be shown, see~\cite{tseng1993dbca}, that
 $\left(-p_i^{(t)}\right)$ and $g_i^{(t)}$ are equivalent in the following sense: Whenever $j(t)=i$, both $\left(-p_i^{(t)}\right)$ and $g_i^{(t)}$ are subgradients of $f_i$ at $x_i^{(t)}$ such that the condition for optimality, \eqref{eq:optimality}, holds.

Finally, note that, by the strong convexity of $f_0$, $x_0 = \nabla f_0^* \left(\sum_{i=1}^n p_i\right)$. Also, $x_0^{(t)} = x_i^{(t)}$ whenever $j(t)=i$, and $x_i^{(t)} = x_i^{(t-1)}$ otherwise. Formally, each $x_i^{(t)}$ is in the sigma-algebra generated by the random variables $\left\{j(1),\ldots,j(t); \mathbf{p}^{(0)},\ldots,\mathbf{p}^{(t)}\right\}$.

\section{Related work}\label{sec:related}

The centralized dual block coordinate ascent method has been thoroughly studied in the literature. We begin this section by clarifying how our contributions differ from previous work, namely, how our set of assumptions prevents us from using previous results to study StochaLM. Later in this section, we present an overview of distributed optimization methods, with special focus on token-based methods.

\subsection{Centralized dual coordinate ascent}

A centralized dual block coordinate ascent (DBCA) method which bears strong similarities with our proposed method is studied in~\cite{tseng1993dbca}. This work proves convergence of DBCA for a more general class of problems than problem~\eqref{prob:main}, in the following sense: The primal and dual sequences generated by DBCA are bounded, and their accumulation points are optimal.

What sets our present work apart from \cite{tseng1993dbca} is the set of assumptions regarding how the $j(t)$ are sampled: The proof of convergence of \cite{tseng1993dbca} assumes that the $j(t)$ are sampled from $\{1,\ldots,n\}$ in an essentially cyclic manner. (Essentially cyclic order means that there exists a finite $T\geq 1$ such that, for all $t$, all the indexes in $\{1,\ldots,n\}$ are selected at least once between $t$ and $(t+T-1)$.) Our assumption that the $j(t)$ follows the state of an ergodic Markov chain breaks this essentially cyclic assumption, thus prevents us from simply using the analysis of~\cite{tseng1993dbca}.

Another version of the Stochastic Dual Coordinate Ascent (SDCA) method is studied in~\cite{shwartz2013sdca}. This works proves convergence results on the expected value of the duality gap when $f_0$ is a quadratic function, and the $f_i$ are either Lipschitz, Lipschitz-smooth, or almost everywhere smooth. We allow for a more generic, strongly-convex $f_0$, and make no assumptions regarding the cost's smoothness. Additionally, as far as we are aware, the strategies for picking $j(t)$ studied in~\cite{shwartz2013sdca} assume either that the $j(t)$ are sampled independently, or that the $j(t)$ are sampled from permutations of the set $\{1,\ldots,n\}$. The latter is a special case of the essentially cyclic assumption proposed in~\cite{tseng1993dbca}, and is therefore not applicable to our case. The former, while having practical interest---as it includes the independent, identically distributed sampling case---does not include the case when the $j(t)$ are sampled from the state of a Markov chain, which is key to allowing StochaLM to be used in a distributed setup. As such, we cannot borrow the analysis presented in~\cite{shwartz2013sdca} to study our method.

A proximal version of SDCA is presented in~\cite{shwartz2016proxsdca}. The updates of this method differ from those of~\cite{shwartz2013sdca} and StochaLM. Its convergence analysis assumes that the cost function components, $f_i$ are either Lipschitz or Lipschitz-smooth. Again, our method does not make such requirements.

The adaptive SDCA method (AdaSDCA) presented in~\cite{csiba2015adasdca} iteratively updates the scheme for picking the $j(t)$ over time in order to accelerate convergence, using knowledge regarding the global Lipschitz constant. We assume we have no control over the scheme governing the sampling the $j(t)$. Furthermore, our method does not require Lipschitz-smoothness, nor does it require knowledge regarding the total cost; each agent needs, at most, information regarding its private cost and $f_0$, and information regarding its immediate neighbourhood.

Quartz~\cite{qu2015quartz} is a method that requires independent sampling of the dual block components. Quartz has one hyperparameter, $\theta$, which cannot be chosen freely. Its adequate value depends on a number of properties regarding the total cost and the distribution of $j(t)$. Our method is more local, in the sense that any parameter that an agent needs to use requires, at most, information regarding its private cost and $f_0$.

The Random Accelerated Coordinate Descent~\cite{necoara2019dualgrad} performs, at each round, a number gradient steps along a (dual) coordinate block, rather than solving a subproblem exactly. This method requires the cost to be Lipschitz smooth.

None of these previous works discuss a distributed implementation of the dual block coordinate ascent method. Furthermore, the proofs presented in these works do not directly extend to our distributed scenario.


\subsection{Distributed optimization}

The stochastic gradient descent (SGD) method proposed in~\cite{johansson2010token} performs, at round $t$, a descent step using a subgradient of $f_{j(t)}$, where $j(t)$ follows the state of an ergodic Markov Chain with uniform stationary distribution. The authors of~\cite{johansson2010token} provide a strategy to locally design transition probabilities which are compatible with the underlying network, and which have an uniform stationary distribution. Following this strategy, each agent in the network can compute its ``outgoing'' probabilities using knowledge of its immediate neighbourhood, without having to know the network's global topology. This method uses a pre-determined global sequence of stepsizes. Our method, in contrast, does not have such global hyperparameters.

The uniform stationary distribution ensures that the token does not visit highly-connected nodes more often than less-connected ones. This bias is addressed in~\cite{ayache2021token}; in this version of SGD, an agent may choose not to pass the token to a neighbour, and perform an additional gradient step instead.

ADMM has also been adapted to token-based communications: Walkman~\cite{mao2020walkman} can solve problems such as~\eqref{prob:main}, even if the $f_i$ are slightly non-convex, as long as they are Lipschitz-smooth. There are two versions of Walkman, the proximal-step version and the gradient-step version, which aim to minimize, respectively, the number of communications and the complexity of internal computations. Walkman as one global hyperparameter which has be tuned using knowledge on the global Lipschitz constant. In contrast, our method does not have global hyperparameters, and does not require the cost components to be Lipschitz-smooth.

Walkman, like SGD, tends to pick highly-connected agents more often than others. In~\cite{ye2020iadmm}, the authors propose that the token follows a Hamiltonian cycle (a cycle that visits all the agents in the network exactly once) to fix this bias. Finding a Hamiltonian path is an NP-complete problem~\cite{karp1972, garey1990}. Finally, a multi-token version of Walkman is presented in~\cite{ye2020padmm}.

We will compare our method with the token-based SGD proposed in~\cite{johansson2010token} and with Walkman~\cite{mao2020walkman} in Section~\ref{sec:numerical}.

Both SGD and ADMM can be used in distributed optimization without a token, see~\cite{nedic2009dsg,feng2014broadcastsgd, arablouei2017sgdbias,xu2018sgdbadagents,barani2021diffusionsgd} and~\cite{boyd2011admm,wei2012admm,wei2013okadmm,iutzeler2013randadmm,zhang2019asynchadmmml}, respectively. Other methods which don't use a token include, but are certainly not limited to, EXTRA~\cite{shi2015extra}, DIGing~\cite{nedic2017digging}, ADD-OPT~\cite{xi2018addopt}, Nesterov methods~\cite{jakovetic2014fastdist,xin2019abn}, augmented Lagrangean methods~\cite{jakovetic2011alm,lee2017alm,zhang2018alm}, variance reduction and average gradient methods~\cite{xin2020svrg,hu2021svrgsag}, and others~\cite{chen2012fastproxgrad,shi2015pgextra,cui2020momentum,barazandeh2021adam,peng2021byzantine}.

\section{Convergence results}\label{sec:stochalm_proof}

Because StochaLM makes use of a token whose movement is described by the state of an ergodic Markov Chain, the essentially cyclic and independence assumptions, made in~\cite{tseng1993dbca} and~\cite{shwartz2013sdca} respectively, are not satisfied. The proofs of these works do not, therefore, generalize to our case. In this section, we present the proof we designed specifically for StochaLM.

We begin by stating our assumptions:

\paragraph*{Assumption 0} The underlying graph of the network is \emph{fully} connected.

\vspace{0.3cm}
This assumption is made explicit since, in distributed applications, one often assumes the network is connected, but not fully so. While we do not have, at the moment, results for connected, but not fully connected, networks, our numerical experiments show that our algorithm can converge to the solution even when the network is not fully connected. As such, it is possible that our proof can be extended to those cases.

\paragraph*{Assumption 1} The $j(t)$, $t=1,2,\ldots,$ follow the state of an aperiodic, irreducible, time-invariant Markov chain with $n$ states, where state $j$ corresponds to the token being on agent $j$. The transition matrix of the Markov chain, $P:=\{P_{ij}\}$, where $P_{ij} = \Prob(j(t) = j \mid j(t-1) = i)$, only allows transitions between connected agents; this means that $P_{ij}>0$ iff $j\in \mathcal{N}_i$. Additionally, $P$ is row-stochastic, meaning that $\sum_{j=1}^n P_{ij} = \sum_{j\in\mathcal{N}_i} P_{ij} = 1$.

\vspace{0.3cm}
Since $n$ is finite, there exists some $P_->0$ such that $P_{ij}\geq P_-$ whenever $P_{ij}>0$. (Under assumption~0, $P_{ij}\geq P_- > 0$ for all $i,j$.) For aperiodic and irreducible Markov chains, the Perron-Frobenius theory (see \cite[Chapter~8]{horn1990matrixanalysis}, for example) guarantees that there exists a unique stationary distribution vector, $\pi>0$, such that $\pi^\top = \pi^\top P$, and such that $\Prob(j(t) = i) \rightarrow \pi_i$ as $t\rightarrow \infty$. An important consequence of this is that all agents are selected infinitely often with probability one. Also, because $x^{(t)}=x^{(t+1)}$ if $j(t)=j(t+1)$, we can assume, without loss of generality, that $j(t)\neq j(t+1)$ for all $t$.

\paragraph{Assumption 2} $f_i:\reals^p\rightarrow \reals\cup\{+\infty\}$ is closed and convex, for $i=0,1,\ldots,n$.

\paragraph{Assumption 3}$f_i$ has closed domain, $\dom f_i=\{x\mid f_i(x)<+\infty\}$, for $i=0,1,\ldots,n$.

\vspace{0.3cm}
Assumptions 2 and 3 are satisfied when all the cost components are real-valued, that is, $\dom f_i = \reals^p$ for $i=0,1,\ldots,n$. Assumptions 2 and 3 are also satisfied when $f_i = h_i + \iota_{C_i}$, where $h_i:\reals^p \rightarrow\reals$ is a real-valued, convex function ($\dom h_i = \reals^p$), and $\iota_{C_i}$ is the indicator of a closed, convex set, $C_i \subset \reals^p$. This allows one to embed implicit constraints in each $f_i$.

\paragraph{Assumption 4} For $i=0,1,\ldots,n$, $f_i$ is strongly convex for some (possibly unknown) $\mu_i>0$, that is, $f_i - \frac{\mu_i}{2}\|\cdot\|^2$ is convex.

\vspace{0.3cm}
Assumption 4 is not necessarily more restrictive than having only $f_0$ be strongly convex: Given that $f_0$ is strongly convex, we can redefine the primal cost components as $f^\epsilon_0 := (1-\epsilon)f_0$ and $f^\epsilon_i := f_i + \frac{\epsilon}{n}f_0$, for $i=1,\ldots,n$, and for some $\epsilon\in (0,1)$, thus ensuring that all the components are strongly convex. This redefinition is harmless in most applications, since $f_0$ plays the role of a data independent regularizer that is known by all the nodes.

\paragraph{Assumption 5} (Slater's condition) There exists a point, $\bar{x}$, such that $\bar{x}\in \bigcap_{i=0}^n \rint\dom f_i$. 

\vspace{0.3cm}
This last assumption trivially implies that there exists $\bar{\mathbf{x}}=(\bar{x}_0,\bar{x}_1,\ldots,\bar{x}_n)$ such that $\mathcal{A}\bar{\mathbf{x}}=0$ and $\bar{\mathbf{x}}\in\rint \dom f$ (matrix $\mathcal{A}$ is defined in \eqref{eq:mathcal_A}). This has an important implication on the properties of the dual cost. Consider the so-called inf-image function of $f$ under $\mathcal{A}$, defined as $(\mathcal{A}f)(\mathbf{x}) := \inf_{\mathbf{y}} \{f(\mathbf{y}) \mid \mathcal{A}\mathbf{y}=\mathbf{x}\}$, with the usual convention that $\inf \emptyset = +\infty$. It can be shown, using Sion's Minimax Theorem~\cite{sion1958minimax}, that $h^*(\mathbf{x})=(\mathcal{A}f)(\mathbf{x})$, where $h$ is the symmetric of the dual cost, defined in~\eqref{eq:main_dual}. This fact, combined with~\cite[Proposition~III.2.1.1]{urruty1993i}, and with the (easily provable) fact that $\dom(\mathcal{A}f) = \mathcal{A}(\dom f)$, means that, under assumption~5, $0\in\rint\dom (\mathcal{A}f)$, or $0\in\rint\dom h^*$. This, in turn, implies is that $h$ is assymptotically well-behaved (a.w.b.), see~\cite[Theorem 4.2.3, p.~131]{teboulle2003awb}: A function $h$ is a.w.b. if any sequence $\left\{ \mathbf{p}^{(t)} \right\}_{t=0}^\infty$ which satisfies $\nabla h\left(\mathbf{p}^{(t)}\right)\rightarrow 0$ also satisfies $h\left(\mathbf{p}^{(t)}\right) \rightarrow \inf h,$ see~\cite[Definitions~4.2.1 and~4.2.2, p.~125]{teboulle2003awb}.

\vspace{0.3cm}
We begin by showing that the primal sequences are bounded.

\begin{theorem}
\label{theo:stochalm_bounded}
Under assumptions 1-5, the primal sequences $x^{(t)}$ and $\mathbf{x}^{(t)} = \left(x_0^{(t)},\ldots,x_n^{(t)}\right)$, $t=1,2,\ldots$, are bounded with probability one.
\end{theorem}

\begin{proof}
Let us define $f_*^{(t)}:= \inf_x f^{(t)}(x)$. Using the fact that $x^{(t)}=x_0^{(t)}=x_{j(t)}^{(t)}$, and that $x_i^{(t)}=x_i^{(t-1)}$, $g_i^{(t)}=g_i^{(t-1)}$ for $i\neq j(t)$, one finds, through a series of simple manipulations, that
\[
f_*^{(t)}=f_0\left(x_0^{(t)}\right) + \sum_{i=1}^n \left\lbrace f_i\left(x_i^{(t)}\right) + \left\langle g_i^{(t)}, x_0^{(t)} - x_i^{(t)}\right\rangle \right\rbrace.
\]
Next, we show that $f_*^{(t)} \leq f_*^{(t+1)}$ for all $t$. Define
\[
f^{\left(t+\frac{1}{2}\right)}(x)=f_0(x) + \hat{f}_{j(t)}^{(t)}(x) + \sum_{i\neq j(t)}  \hat{f}_i^{(t-1)}(x),
\]
where, following definition~\eqref{eq:f_hat}, $\hat{f}_{j(t)}^{(t)}(x):= f_{j(t)}\left(x_{j(t)}^{(t)}\right) + \left\langle g_{j(t)}^{(t)}, x - x_{j(t)}^{(t)}\right\rangle$.

Since the conditions for optimality for $f^{(t)}$ and $f^{\left(t+\frac{1}{2}\right)}$ are the same, see \eqref{eq:optimality}, we must have $f_*^{(t)}=f_*^{\left(t+\frac{1}{2}\right)}$. In the next iteration, we have that
\[
f^{(t+1)}(x)=f_0(x) + f_{j(t+1)}(x) + \sum_{i\neq j(t+1)}  \hat{f}_i^{(t)}(x).
\]
By convexity of the $f_i$, we have $\hat{f}_{j(t+1)}(x) \leq f_{j(t+1)}(x)$. Additionally, we have, for all $i\neq j(t), j(t+1)$, that $\hat{f}_i^{(t)}(x)= \hat{f}_i^{(t-1)}(x)$, since $x_i$ and $g_i$ did not change in the meantime. One concludes that  $f_*^{\left(t+\frac{1}{2}\right)} \leq f_*^{(t+1)}$, from which follows immediately that $f_*^{(t)} \leq f_*^{(t+1)}$.

The convexity of the $f_k$ also implies that $f^{(t)} \leq F$ for all $t$, and, thus, that $f_*^{(t)} \leq F_*$. We conclude that $\left\{f_*^{(t)}\right\}_{t=1}^\infty$ is a non-decreasing sequence bounded from above by $F_*$, and must, therefore, converge to some $f_*^\infty \leq F_*$.

Since $f_0$ is strongly convex, we have that $\frac{\mu}{2}\left\|x^{(t)}-x^*\right\|^2 \leq F(x^*) - f_*^{(t)}$, and, since the $f_*^{(t)}$ are bounded, we conclude that $x^{(t)}$ is bounded. It immediately follows that $x_0^{(t)}\equiv x^{(t)}$ is also bounded.

Finally, pick any $i\in\{1,\ldots,n\}$. Because $x_i^{(t)}=x_0^{(t)}$ if $j(t)=i$, and $x_i^{(t)}=x_i^{(t-1)}$ otherwise, one can show that $x_i^{(t)}$ is bounded as well. Since this holds for all $i$, we must have that $\mathbf{x}^{(t)}$ is bounded.
\end{proof}

Our next three results concern the dual sequence generated by StochaLM:

\begin{theorem}\label{theo:dual_gradient}
Under assumptions 0-5, the dual iterates generated by StochaLM, $\mathbf{p}^{(t)}$, are such that, with probability one, $\nabla h\left(\mathbf{p}^{(t)}\right)\rightarrow 0$.
\end{theorem}

\begin{proof}
Assumption~4 implies that each $f_i^*$ is Lipschitz-smooth with, say, constant $L^*_i$, meaning that $h$ is Lipschitz-smooth for some constant $L^*$. Recall that
\[
p_{j(t)}^{(t)}\in
\arg\min_p\ h\left(p^{(t-1)}_1,\ldots,p^{(t-1)}_{j(t)-1},p,p^{(t-1)}_{j(t)+1},\ldots,p^{(t-1)}_n\right),
\]
and let $h_i^\dagger(\mathbf{p}_{-i}) =\min_p\ h(p_1,\ldots,p_{i-1},p,p_{i+1},\ldots,p_n)$.

An $L$-smooth function $\phi$ satisfies $\phi(y)\leq \phi(z) + \langle \nabla \phi(z), y-z \rangle + \frac{L}{2}\|y-z\|^2$; in particular, this holds for $y=z-\frac{1}{L}\nabla\phi(z)$:
\begin{equation}\label{eq:lipschitz_property}
\inf \phi = \phi^* \leq \phi\left(z-\frac{1}{L}\nabla\phi(z)\right) \leq \phi(z) - \frac{1}{2L}\|\nabla \phi(z)\|^2.
\end{equation}

Our goal is to compute the expected value $\ExpFt{h\left(\mathbf{p}^{(t+1)}\right)}$, where $\mathcal{F}_t$ is the sigma-algebra generated by $\left\{j(1),\ldots,j(t); \mathbf{p}^{(0)},\ldots,\mathbf{p}^{(t)}\right\}$.\footnote{We are assuming that $\mathbf{p}^{(t)}$ is a measurable function. This will be true unless the $f_i$ are extremely pathological.} Notice that
\begin{align*}
\ExpFt{h(\mathbf{p}^{(t+1)})}
&= \sum_{i=1}^n P_{j(t)i} h_i^\dagger\left(\mathbf{p}_{-i}^{(t)}\right) \\
&\overset{(a)}{\leq} \sum_{i=1}^n P_{j(t)i} \left(
h\left(\mathbf{p}^{(t)}\right) - \frac{1}{2L^*} \left\|\nabla_i h\left(\mathbf{p}^{(t)}\right)\right\|^2
\right) \\
&\overset{(b)}{\leq} h\left(\mathbf{p}^{(t)}\right) - \frac{P_-}{2L^*} \sum_{i=1}^n \left\|\nabla_i h\left(\mathbf{p}^{(t)}\right)\right\|^2 \\
&= h\left(\mathbf{p}^{(t)}\right) - \frac{P_-}{2L^*} \left\|\nabla h\left(\mathbf{p}^{(t)}\right)\right\|^2.
\end{align*} 
We used bound~\eqref{eq:lipschitz_property} and the fact that $P_{j(t)i}> P_-$ in $(a)$ and $(b)$, respectively. Applying the Robbins-Siegmund Lemma~\cite{robbins1971rslemma} to the resulting inequality yields that $h(\mathbf{p}^{(t)})\rightarrow h^\infty$, for some measurable $h^\infty$, and $\sum_{t=1}^\infty \left\|\nabla h\left(\mathbf{p}^{(t)}\right)\right\|^2 < \infty$, all with probability one. The latter means that $\nabla h\left(\mathbf{p}^{(t)}\right)\rightarrow 0$ almost surely.
\end{proof}

The next result is a direct consequence of the previous theorem, together with the fact that $h$ is a.w.b.:

\begin{theorem}\label{theo:dual_cost}
Under assumptions 0-5, we have that $h\left(\mathbf{p}^{(t)}\right) \rightarrow \inf h$, with probability one.
\end{theorem}

We can now prove our main result concerning the dual sequence that is generated by StochaLM:

\begin{theorem}\label{theo:dual_optimal}
Under assumptions 0-5, and with probability one, the dual iterates generated by StochaLM, $\mathbf{p}^{(t)}$, $t=1,2,\ldots$, are bounded. Furthermore, all accumulation points of $\mathbf{p}^{(t)}$ are optimal.
\end{theorem}

\begin{proof}
Consider a vanishing (and minimizing) sequence, $\mathbf{p}^{(t)}$. Let $\mathbf{q}^{(t)}\in \nabla h(\mathbf{p}^{(t)})$, with a slight abuse of notation. Then, per~\cite[Proposition~X.1.4.3]{urruty1993i}, $\mathbf{p}^{(t)}\in\partial h^*(\mathbf{q}^{(t)})$. Since $\mathbf{q}^{(t)}\rightarrow 0$, by Theorem~\ref{theo:dual_gradient}, and $0\in\rint\dom h^*$, by Assumption~5, we have, by~\cite[Proposition~VI.6.2.2]{urruty1993i}\footnote{The cited result is for real-valued functions, but its proof can be adapted to extend the result for extended-value functions.}, that the $\mathbf{p}^{(t)}$ are bounded.

Now pick an accumulation point of $\mathbf{p}^{(t)}$, $\mathbf{p}^*$; such a point is guaranteed to exist because $\mathbf{p}^{(t)}$ is bounded. By closedness of $h$, and by Theorem~\ref{theo:dual_cost}, we have that $h(\mathbf{p}^*) \leq \lim \inf h\left(\mathbf{p}^{(t)}\right) = \inf h$, so $\mathbf{p}^*$ must be optimal.
\end{proof}

This concludes our results in terms of the dual sequence. We now show what they imply in terms of the primal sequence, $\mathbf{x}^{(t)}$.

\begin{theorem}\label{theo:primal}
Under assumptions 0-5, and with probability one, all the accumulation points of the primal iterates generated by StochaLM, $\mathbf{x}^{(t)}$, $t=1,2,\ldots$, are optimal.
\end{theorem}

\begin{proof}
The KKT conditions for optimality of problem~\eqref{eq:main_dist_vector} are
\begin{align}
\mathcal{A}^\top\mathbf{p}^* &\in \partial f(\mathbf{x}^*) \label{eq:kkt1} \\
\mathcal{A}\mathbf{x}^* &=0. \label{eq:kkt2}
\end{align}
Note that \eqref{eq:kkt2} implies that $\mathbf{x}^*$ is of the form $\mathbf{x}^*:= (x^*,x^*,\ldots, x^*)\in\reals^{(n+1)p}$. Our goal is to show that any accumulation point of $\mathbf{x}^{(t)}$ satisfies~\eqref{eq:kkt1} and~\eqref{eq:kkt2}.

Consider a subsequence $\left\{t(\tau)\right\}_{\tau=1}^\infty$ such that $\mathbf{x}^{(t(\tau))}\rightarrow \mathbf{x}^\infty$. Because the $\mathbf{p}^{(t)}$ are bounded, there exists a sub-subsequence of $\left\{t(\tau)\right\}$, $\left\{t(\tau(m))\right\}_{m=1}^\infty$, such that $\mathbf{p}^{(t(\tau(m)))}\rightarrow \mathbf{p}^*$, where $\mathbf{p}^*$ is optimal as per Theorem~\ref{theo:dual_optimal}. To avoid overcluttering the notation, we can assume, without loss of generality, that $\left\{t(\tau)\right\}_{\tau=1}^\infty$ itself is that subsequence.

By assumption~4, both $x_0$ and the $x_i$ are uniquely determined given some value of $\mathbf{p}$ (cf. Section~\ref{sec:stochalm_dual}); in particular, $\mathbf{x}^{(t)}= \nabla f^*\left(\mathcal{A}^\top\mathbf{p}^{(t)}\right)$. Taking the limit $\tau\rightarrow\infty$ in $\mathbf{x}^{(t(\tau))}= \nabla f^*\left(\mathcal{A}^\top\mathbf{p}^{(t(\tau))}\right)$ yields $\mathbf{x}^\infty= \nabla f^*\left(\mathcal{A}^\top\mathbf{p}^*\right)$, which, by~\cite[Proposition~X.1.4.3]{urruty1993i}, is~\eqref{eq:kkt1}.

Recall, now, the definition $\mathbf{q}^{(t)} = \nabla h\left(\mathbf{p}^{(t)}\right)$ from previous proofs. Because $f$ is strongly convex, $\dom f^*=\reals^{(n+1)p}$, allowing us to use~\cite[Lemma X.2.2.2]{urruty1993ii}, together with the chain rule, to find that $\mathbf{q}^{(t)}= \mathcal{A} \mathbf{x}^{(t)}$. Taking the limit $\tau\rightarrow\infty$ in $\mathcal{A}\mathbf{x}^{(t(\tau))}=\mathbf{q}^{(t(\tau))}$ yields $\mathcal{A}\mathbf{x}^\infty=0$, which is~\eqref{eq:kkt2}. We thus conclude that $\mathbf{x}^\infty$ is optimal, or $\mathbf{x}^\infty = \mathbf{x}^*$.
\end{proof}

Finally, we combine the previous result with Theorem~\ref{theo:stochalm_bounded}:

\begin{theorem}\label{theo:primal_converges}
Suppose assumptions 0-5 hold. Then, with probability one, the primal iterates generated by StochaLM, $\mathbf{x}^{(t)}$, $t=1,2,\ldots$, are such that $x_i^{(t)}\rightarrow x^*$, for $i=0,1,\ldots,n$.
\end{theorem}

\begin{proof}
By the boundedness of the $\mathbf{x}^{(t)}$ (Theorem~\ref{theo:stochalm_bounded}), at least one accumulation point exists. Theorem~\ref{theo:primal} states that any accumulation point must be optimal with probability one. Assumption~4 implies there is only one optimal point, $\mathbf{x}^* = (x^*,\ldots,x^*)$. Therefore, all accumulations points of $\mathbf{x}^{(t)}$ are equal to $\mathbf{x}^*$. This, together with the fact that the $\mathbf{x}^{(t)}$ are bounded, implies that $\mathbf{x}^{(t)}\rightarrow\mathbf{x}^*$ with probability one.
\end{proof}

With this result, we conclude our theoretical analysis of StochaLM. We now move on to the numerical results.

\section{Numerical results}\label{sec:numerical}

In this section, we compare the distributed version of StochaLM with the token-based SGD method proposed in~\cite{johansson2010token}, and with Walkman~\cite{mao2020walkman}. To facilitate the reproduction of our results, we provide additional implementation details over the next paragraphs.

We compare all three methods in a least squares with elastic net regularization~\cite{zou2005elasticnet} problem, with $f_0(x) = \lambda_1\|x\|_1 + \frac{\lambda_2}{2}\|x\|_2^2$, and $f_i(x) = \frac{1}{2n}\|A_i x - y_i\|_2^2$, for $i=1,\ldots,n$.

The dataset of agent $i$ is composed of a feature matrix, $A_i\in\reals^{m\times p}$, and an observed value vector, $y_i\in\reals^m$, which are generated as follows: First, generate a vector $x^{\mathrm{true}}\in\reals^p$ by sampling from a uniform distribution $\mathcal{U}(-1,1)$. Then, generate each $A_i$ by sampling from $\mathcal{U}(-1,1)$ as well. Finally, generate the observed data as $y_i=A_i x^{\mathrm{true}} + v_i$, where $v_i$ is random noise generated from a zero-mean Gaussian distribution with a relatively small variance. Notice that, in general, $x^*\neq x^{\mathrm{true}}$.

While we have verified that, in practice, it is not necessary that the $f_i$ are strongly convex, the results for StochaLM reported here use the re-defined functions $f_0^\epsilon$ and $f_i^\epsilon$ presented in Section~\ref{sec:stochalm_proof}, with $\epsilon=0.01$. Furthermore, since SGD does not handle a separate regularizer function, we integrate $f_0$ within each agent's personal cost for this method.

We also make note of the fact that, while the proofs of convergence for StochaLM require the underlying network to be fully connected, the results in this section are for a connected, but not fully connected, network. The transition probabilities of the Markov chain are compatible with the underlying network, and they are computed using the local strategy proposed in~\cite{johansson2010token}.

We test two versions of SGD, one with a constant stepsize, and one where the stepsizes shrink with $1/t$. For Walkman, we tested several values for the hyperparamenter, $\beta$, and picked the one that yielded the fastest convergence. We found this value to be $\beta =1.01 \beta_\mathrm{min}$, where $\beta_\mathrm{min}$ is the minimum value that is theoretically allowed; using $\beta=\beta_\mathrm{min}$ caused numerical instablilities. We included both proximal and gradient variants of Walkman in our benchmark.

The centralized problem \eqref{prob:main}, and the subproblems of StochaLM and Walkman, are solved using CVX~\cite{cvx,grant2008cvx}. We focus on the evolution of the relative error of the token variable, $e^{(t)}:=\frac{\|x_0^{(t)}-x^*\|_2}{\|x^*\|_2},$ averaged across several Monte Carlo runs. In each Monte Carlo run, a different sequence of $j(t)$ is chosen.

\begin{figure*}[t] 
\centering
\begin{subfigure}[b]{0.45\textwidth}
     \centering
     \includegraphics[trim={1.3cm 1.5cm 1.3cm 0.5cm},clip,width=1.0\textwidth]{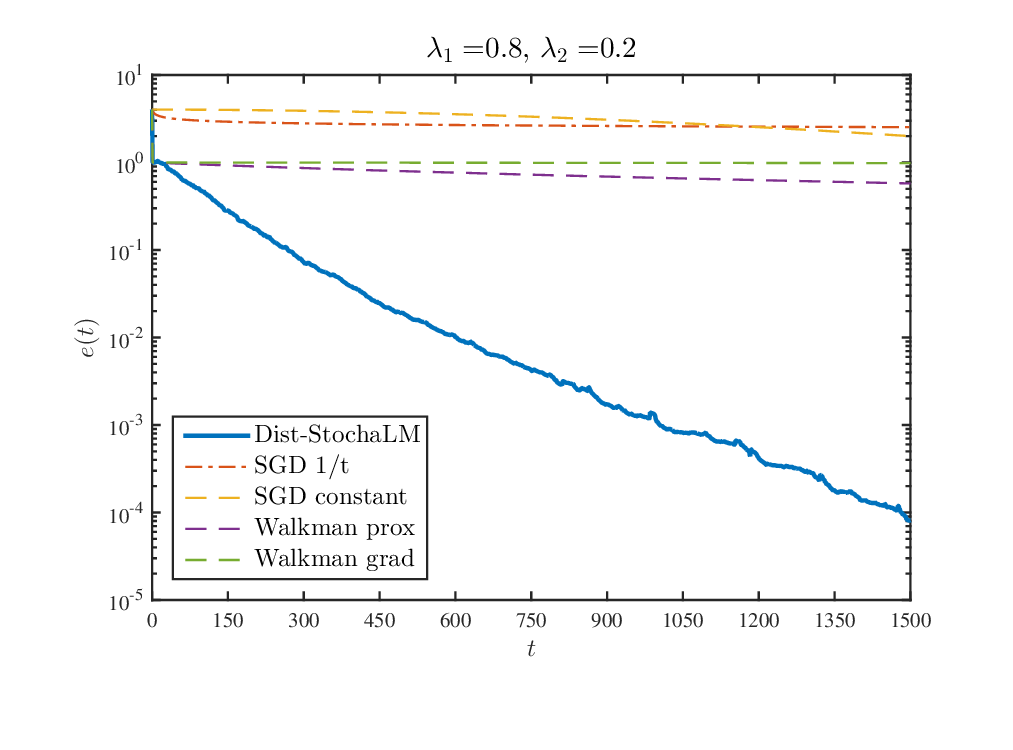}
     \caption{$m=3$}
     \label{fig:central_m3}
\end{subfigure}
\begin{subfigure}[b]{0.45\textwidth}
     \centering
     \includegraphics[trim={1.3cm 1.5cm 1.3cm 0.5cm},clip,width=1.0\textwidth]{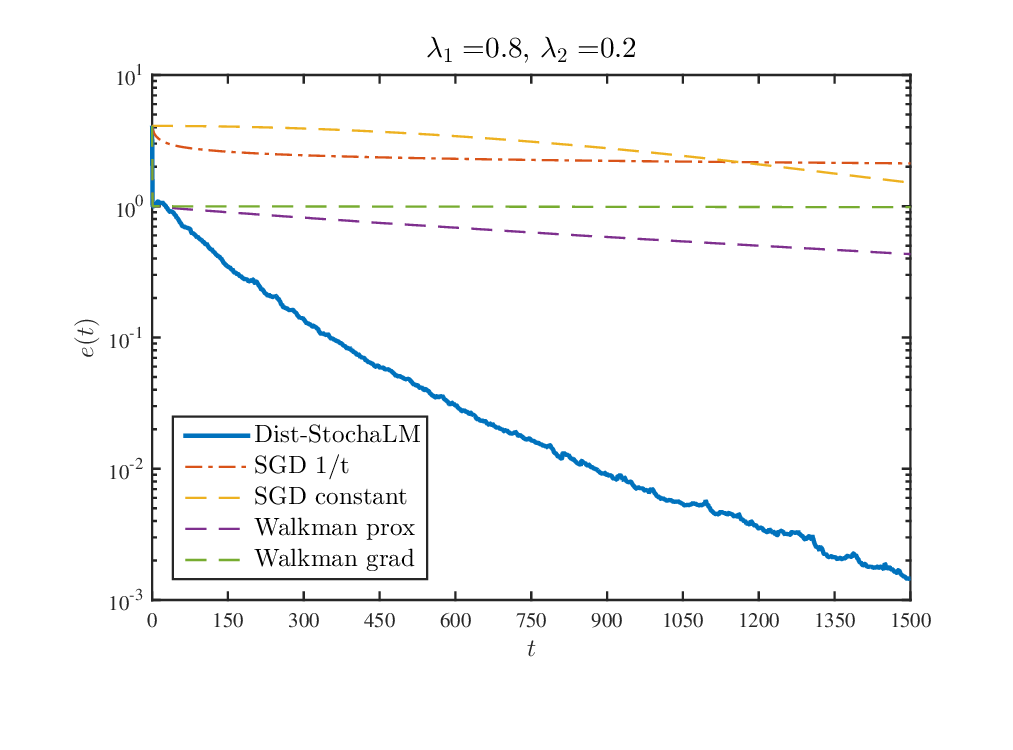}
     \caption{$m=6$}
     \label{fig:central_m6}
\end{subfigure}
\caption{Mean relative error across $20$ Monte Carlo trials on an elastic-net regression experiment with $n=30$ agents, $p=120$ features, and $m$ observations per agent.}
\label{fig:token1}
\end{figure*}

The results for this numerical experiment are shown in Figure~\ref{fig:token1}. SGD is the slowest method, and its convergence speed is strongly affected by the shrinking stepsizes. Both our method and Walkman make a significant ``jump'' on the first iteration; we believe this happens because both these methods perform an exact optimization step (both versions of Walkman contain a proximal step in $f_0$), while SGD takes a single gradient step, at each round. Walkman, perhaps because it was designed to handle some non-convexity, has worse performance than StochaLM in this (strongly) convex problem. We verified that the error of gradient-step version of Walkman is shrinking linearly, although this is not visible in Figure~\ref{fig:token1}.

We also observe that the performance of the three methods changes slightly, depending on whether there are more features ($m=3$) or total observations ($m=6$). In particular, StochaLM becomes slightly slower (reaching a precision a little above $10^{-3}$ instead of $10^{-4}$), and Walkman slightly faster, when $m$ increases. Potentially, Walkman could surpass StochaLM for large enough values of $m$. We did not pursue this matter further, as regularized regression problems of interest typically have fewer total observations than features.

\section{Conclusions}\label{sec:conclusions}

StochaLM is a token-based distributed optimization method with no global hyperparameters. The movement of the token is described by the state of an ergodic Markov chain, and is neither essentially cyclic nor independent over time. As such, theoretical convergence results from previous works do not apply directly to our method. We showed that the method converges to the centralized solution, with probability one, in fully connected networks, and when the cost is strongly convex. We report numerical results which evidence that StochaLM performs better than other token-based methods, even when the latter's hyperparameters are set for optimal performance.


\bibliographystyle{elsarticle-num} 
\bibliography{stochalm,stochalm_state}

\end{document}